\documentclass[preprint]{revtex4-1}

\usepackage{graphicx}
\usepackage{dcolumn}
\usepackage{bm}
\usepackage{color}
\usepackage{gensymb}

\newcommand{\sini}{Si$_3$N$_4$}
\newcommand{\liii}{L$_3$}
\newcommand{\lii}{L$_2$}

\begin{document}

\title{Soft x-ray absorption of thin films detected using substrate luminescence: A performance analysis}

\author{Cinthia Piamonteze}
\email{cinthia.piamonteze@psi.ch}
\affiliation{Swiss Light Source, Paul Scherrer Institut, CH-5232 Villigen PSI, Switzerland}

\author{Yoav William Windsor}\altaffiliation[Present address: ] {Department of Physical Chemistry, Fritz Haber Institute of the Max Planck Society, Faradayweg 4-6, 14195 Berlin, Germany}
\affiliation{Swiss Light Source, Paul Scherrer Institut, CH-5232 Villigen PSI, Switzerland} 

\author{Sridhar R. V. Avula }
\affiliation{Swiss Light Source, Paul Scherrer Institut, CH-5232 Villigen PSI, Switzerland}

\author{Eugenie Kirk}
\affiliation{ Paul Scherrer Institut, 5232 Villigen PSI, Switzerland \\ Laboratory for Mesoscopic Systems, Department of Materials, ETH Zurich, 8093 Zurich, Switzerland} 

\author{Urs Staub}
\affiliation{Swiss Light Source, Paul Scherrer Institut, CH-5232 Villigen PSI, Switzerland}



\date{}

\begin{abstract}
X-ray absorption spectroscopy of thin films is central to a broad range of scientific fields, and is typically detected using indirect techniques. X-ray excited optical luminescence (XEOL) from the sample's substrate is one such detection method, in which the luminescence signal acts as an effective transmission measurement through the film. This detection method has several advantages that make it versatile compared to others, in particular for insulating samples or when a probing depth larger than 10\,nm is required. In this work we present a systematic performance analysis of this method with the aim of providing guidelines for its advantages and pitfalls, enabling a wider use of this method by the thin film community. We compare and quantify the efficiency of XEOL from a range of commonly used substrates. Our measurements demonstrate the equivalence between XEOL and x-ray transmission measurements for  thin films. Moreover, we show the applicability of XEOL to magnetic studies by employing XMCD sum rules with XEOL-generated data. Lastly, we demonstrate that above a certain thickness XEOL shows a saturation-like effect, which can be modelled and corrected for.
\end{abstract}

\maketitle

%

\section{\label{sec:introduction}Introduction}

The study of heterostructures and thin films of complex systems has become very important due to the novel properties that can be obtained at the interface of dissimilar materials or stabilized by strain engineering\cite{Zubko:2011ho}.   The element specific information provided by soft x-ray absorption spectroscopy (XAS) combined with the magnetic and orbital information provided by x-ray magnetic circular dichroism (XMCD) and x-ray linear dichroism (XLD), respectively, made these   techniques widely used in investigation of thin films.

Most materials of interest in this field contain 3d  transition metals or lanthanides, for which large XMCD and XLD contrast appears for L$_{2,3}$ (2p $\rightarrow$ 3d transitions) and M$_{4,5}$ (3$d$ $\rightarrow$ 4$f$ transitions),   respectively.  These edges lie at energies where the x-ray mean free path is on the order of 50\,nm to 100\,nm. This makes transmission measurements for epitaxially grown films and heterostructures challenging. Therefore, indirect detection methods like total electron yield (TEY) or total fluorescence yield (TFY) are commonly used. TEY is a  measurement of the drain current from the sample, which is created to compensate for electrons ejected as a consequence of the x-ray absorption process. TEY  probing depth is typically 5-10\,nm and requires a conducting surface.  TFY measures fluorescence x-ray photons generated due to the photo-absorption effect, and therefore probe to depths of around 100\,nm. Moreover, since photons are measured, the samples do not need to be conducting. However, while spectra acquired from TEY are often used for sum rule analysis of XMCD data, this is not reliable for TFY, as it suffers from self-absorption \cite{Iida:1993en,Troger:1992ud} and often it does not reflect the intrinsic XAS \cite{deGroot:2012kj}. To overcome the limitations of TFY,  the use of inverse partial fluorescence yield can be employed~\cite{Achkar:2011dh,Achkar:2011dh}. In this method one  measures non-resonant x-ray emission by using an energy dispersive fluorescence detector.

X-ray excited optical luminescence (XEOL) is an alternative detection method to obtain XAS of thin films or heterostructures.  X-rays absorbed in an insulator create high energy electrons that produce further conduction electrons and valence holes by electron-electron scattering. Some of these electron-hole pairs radiatively recombine through defect states emitting luminescence\cite{Elango:1994im}. This gives rises to the so called XEOL signal. This technique was initially proposed as a way to detect site specific XAS in the hard x-ray energy range\cite{Bianconi:1978vr}. In recent years XEOL has been introduced as an alternative detection method for measuring soft x-ray transmission through thin films \cite{Kallmayer:2007dz,Vaz:2013dq,Vaz:2012cg}. X-rays transmitted through the  film create  optical luminescence in the substrate. This method has been  applied to collect XAS and XMCD from thin oxide films  \cite{Kallmayer:2007dz,Meinert:2011dv,Cao:2015fd,Green:2015bq,Windsor:2017jk,Aeschlimann:2018jga} and also for spatially resolved measurements using scanning transmission x-ray microscopy (STXM) \cite{Vaz:2013dq,Vaz:2012cg}. Vaz \textit{et al.} \cite{Vaz:2013dq} discuss in detail the origin of the photoluminescence in some substrates. For example, in MgO   the photoluminescence is associated with Cr$^{3+}$ impurities, while in  LAO the photoluminescence is attributed to defects in twin boundaries.

 In this work we explore the strengths and shortcomings of this detection method, which may allow its wider use in the research of strain engineered thin films and heterostructures. The strength of the XEOL method is that it probes the whole thin film thickness and allows measuring insulating  films. For planning XAS measurements by XEOL, it is important to know the x-ray luminescence efficiency of the substrate. For this reason we present here the XEOL efficiency for different substrates. In addition we discuss the application of XMCD sum rules. Finally, we give guidelines for limits on  film thickness required to avoid artifacts in the measurements and we propose a method in how to correct for saturation-like effects in the measurements.

\section{\label{sec:experimental}Experimental}

Co films of 20\,nm thickness were DC-magnetron sputtered in one deposition on nine commonly employed substrates listed in Table \ref{table:efficiency}. For quantifying thickness effects,  Co films of 10, 40 and 80\,nm thickness were sputtered on MgO. Lastly, a reference sample was produced for measuring X-ray transmission: a 20 \,nm thick Co film was sputtered on 100\,nm thick Si$_3$N$_4$. All films were  capped  with 3\,nm or 5\,nm sputtered Pt before breaking vacuum, to avoid oxidation. The film thicknesses were verified by x-ray reflectivity. 

The x-ray absorption measurements were carried out at the X-Treme beamline \cite{Piamonteze:2012jg} in the Swiss Light Source. The XAS spectra were obtained by taking the natural logarithm of the normalized data, following the Beer-Lambert law for linear absorption:
\begin{equation}
    I=I_0 e^{-\mu t} \Rightarrow 
    \mu t = ln (I_0/I)\label{eq_xas}
\end{equation}

In the equation above, $I$ and $I_0$ are the transmitted and incident x-ray intensities, respectively. $\mu$ is the linear absorption coefficient and $t$ is the film thickness. In our case $I$ is the XEOL signal  and $I_0$ is measured in a gold mesh upstream from the sample. To account for the energy dependence of XEOL, we normalize $I^T=I/I_0$ signal measured for the Co/substrate to the $I^T$ of the bare substrate. Finally we take the logarithm to obtain the actual absorption spectrum. The XEOL signal was measured using a UVG100 photodiode  from OptoDiode (http://optodiode.com). The photodiode is mounted directly behind the sample, assuring a large solid angle collection. The incoming x-ray flux, used to create table \ref{table:efficiency} was measured using an AXUV100 photodiode \cite{Kjornrattanawanich:gs}.

XMCD measurements of the Co/substrate films were carried out with a field of 2\,T applied  parallel to the X-ray beam. The films were at 60$^o$ incidence angle (between the film normal and the x-ray beam).

\section{Results and Discussion \label{sec:results}}

\subsection{XEOL efficiency}

\begin{figure}
\begin{center}

\includegraphics[width=0.8\textwidth]{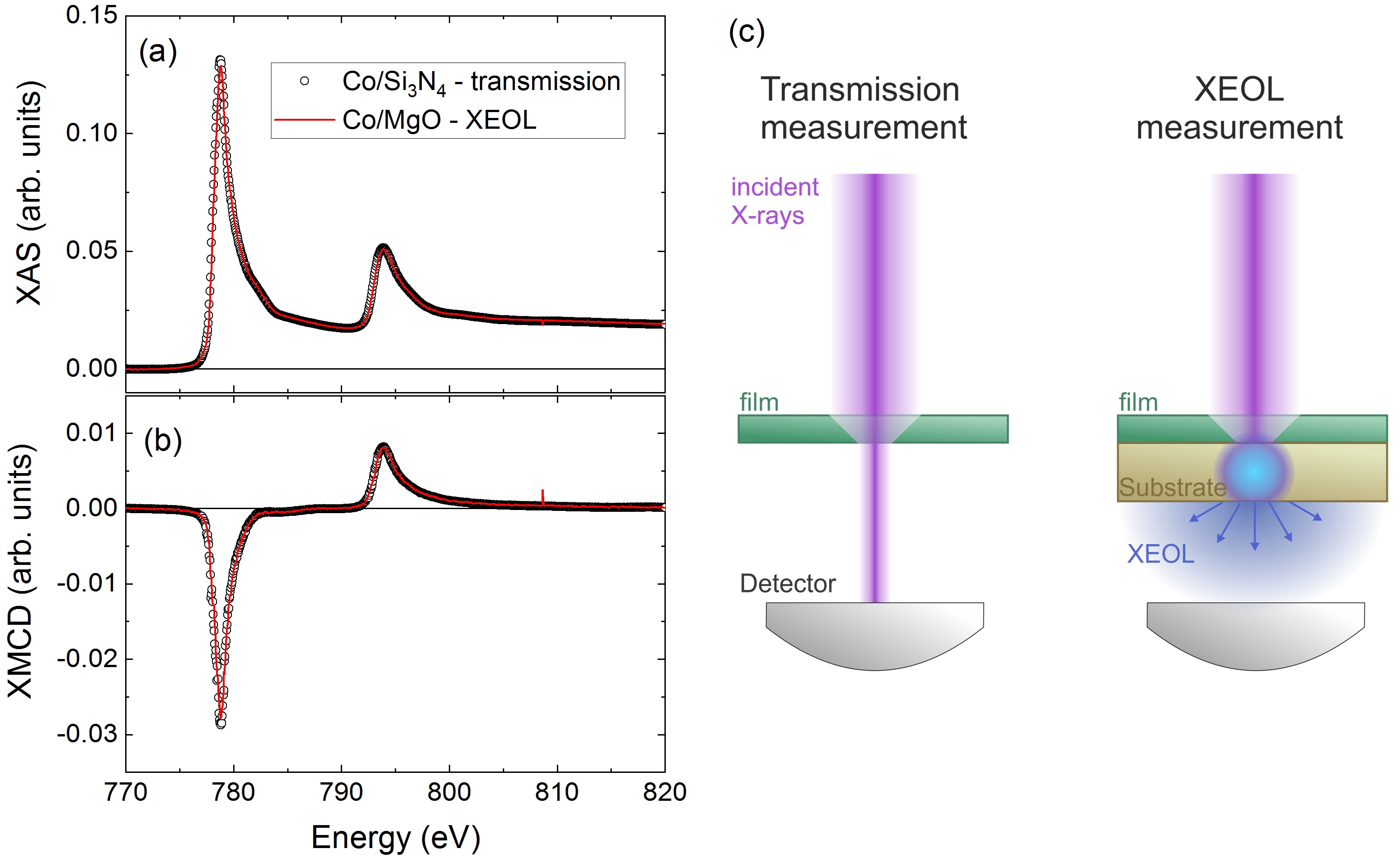}
\caption{\label{fig:transmission} (a) XAS  and (b) XMCD at the Co edge for   Co/Si$_3$N$_4$ measured in transmission (open circles) and  Co/MgO measured using XEOL (line). 
(c) Sketch of measurement setup  for XEOL and transmission measurements.
} 
\end{center}

\end{figure}

We begin by demonstrating the equivalence of XEOL and transmission measurements. Figures \ref{fig:transmission}(a), (b) present a direct comparison between the two measurements, which are sketched in figure \ref{fig:transmission}(c). The figures present XAS and XMCD spectra measured at the Co L$_{3,2}$ edges from 20\,nm thick Co films deposited in parallel on \sini\ and on MgO. The Co/\sini\ data were collected in transmission while the Co/MgO data were collected using XEOL. The excellent agreement between them demonstrates the equivalence of  these detection modes.

\begin{figure}
\begin{center}
\includegraphics[width=0.8\textwidth]{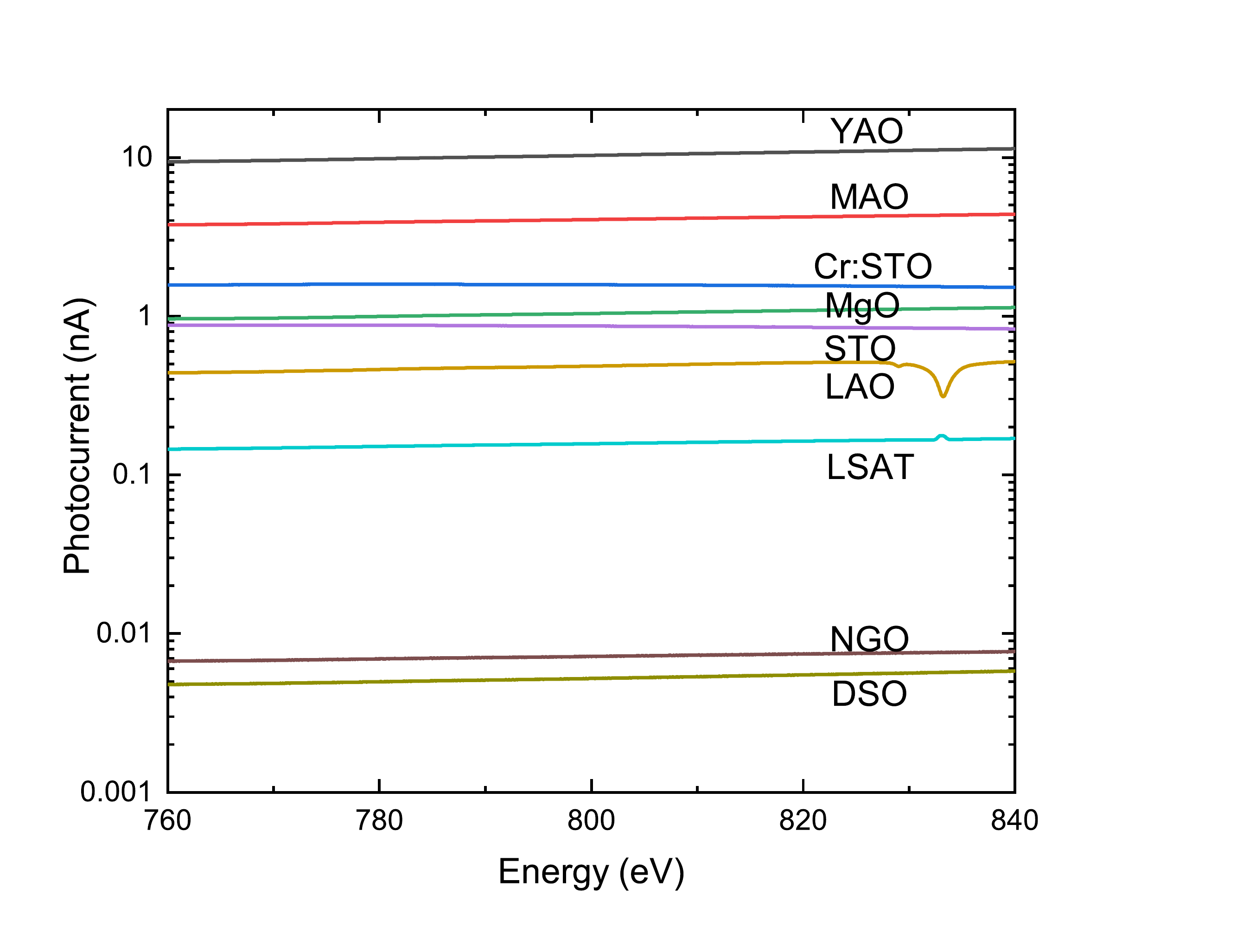}
\caption{\label{fig:substrate} XEOL photocurrent  measured behind substrate in 60$^o$  incidence. The measurements were done at room temperature, except for MAO, STO and Cr:STO which were measured at  100\,K. The normal flux conditions used for almost all substrates was with front-end slit opening of 0.5x0.5\,mm. For DSO and NGO the horizontal front-end slit was further open to 1.0\,mm  and 1.7\,mm, respectively.}
\end{center}
\end{figure}

We now discuss the efficiency of XEOL measurements for different substrates.
 Table \ref{table:efficiency} lists the XEOL efficiency and figure \ref{fig:substrate} presents the XEOL signal of the bare substrates as a function of x-ray energy near the Co L edges. The XEOL efficiency  reflects how many optical luminescence photons are produced per x-ray photon that reaches the substrate. The values on this table were obtained using the photodiode responsivity for 1.94\,eV  (639.5\,nm) radiation. This energy lies between the luminescence maxima measured for LAO and MgO \cite{Vaz:2013dq}. The photodiode used measures an integrated energy signal and its working range is between 300\,nm and 1000\,nm with optimal efficiency at 800\,nm(http://optodiode.com). Notice that efficiencies are temperature dependent and MAO, STO and Cr:STO were measured at  100\,K in order to obtain a measurable photocurrent. At 100\,K the XEOL signal of STO is comparable to MgO at room temperature. Moreover, the addition of Cr impurities increases the XEOL signal from STO by almost a factor of 2. DSO and NGO substrates have a very low efficiency, as shown on table \ref{table:efficiency}. For this reason the data shown in figure \ref{fig:substrate} for DSO and NGO were measured with approximately three and two times higher flux, respectively, than for the other substrates.  Since the luminescence depends on impurities and crystal defects there could be variations depending on the substrate manufacturer. Even so, table \ref{table:efficiency} gives the order of magnitude of the expected signal when designing an  experiment using XEOL as detection mode.

\begin{table}
\begin{center}
\caption{XEOL efficiency measured for bare substrates in grazing incidence.  Measurements at RT except for Cr:STO,  STO and MAO which were measured at 100\,K }
\begin{tabular}{ccc} \hline \hline
Substrate & abbreviation & XEOL   \\
 & & efficiency\\
\hline

YAlO$_3$ & YAO & 2.7e-1\\

MgAl$_2$O$_4$ & MAO & 1.1e-1\\

Cr:SrTiO$_3$ & Cr:STO & 4.5e-2 \\

MgO & MgO  & 2.8e-2 \\

SrTiO$_3$ & STO & 2.5e-2  \\

(LaAlO$_3$)$_{0.3}$-(Sr$_2$AlTaO$_6$)$_{0.7}$ & LSAT & 4.2e-2\\

LaAlO$_3$ & LAO & 1.2e-2 \\

NdGaO$_3$ & NGO & 1.0e-4 \\

DyScO$_3$ & DSO & 6.1e-5 \\ 

\hline  \hline
\end{tabular}
\label{table:efficiency}
\end{center}
\end{table}

\subsection{Measurement speed}

\begin{figure}
\begin{center}
\includegraphics[width=0.8\textwidth]{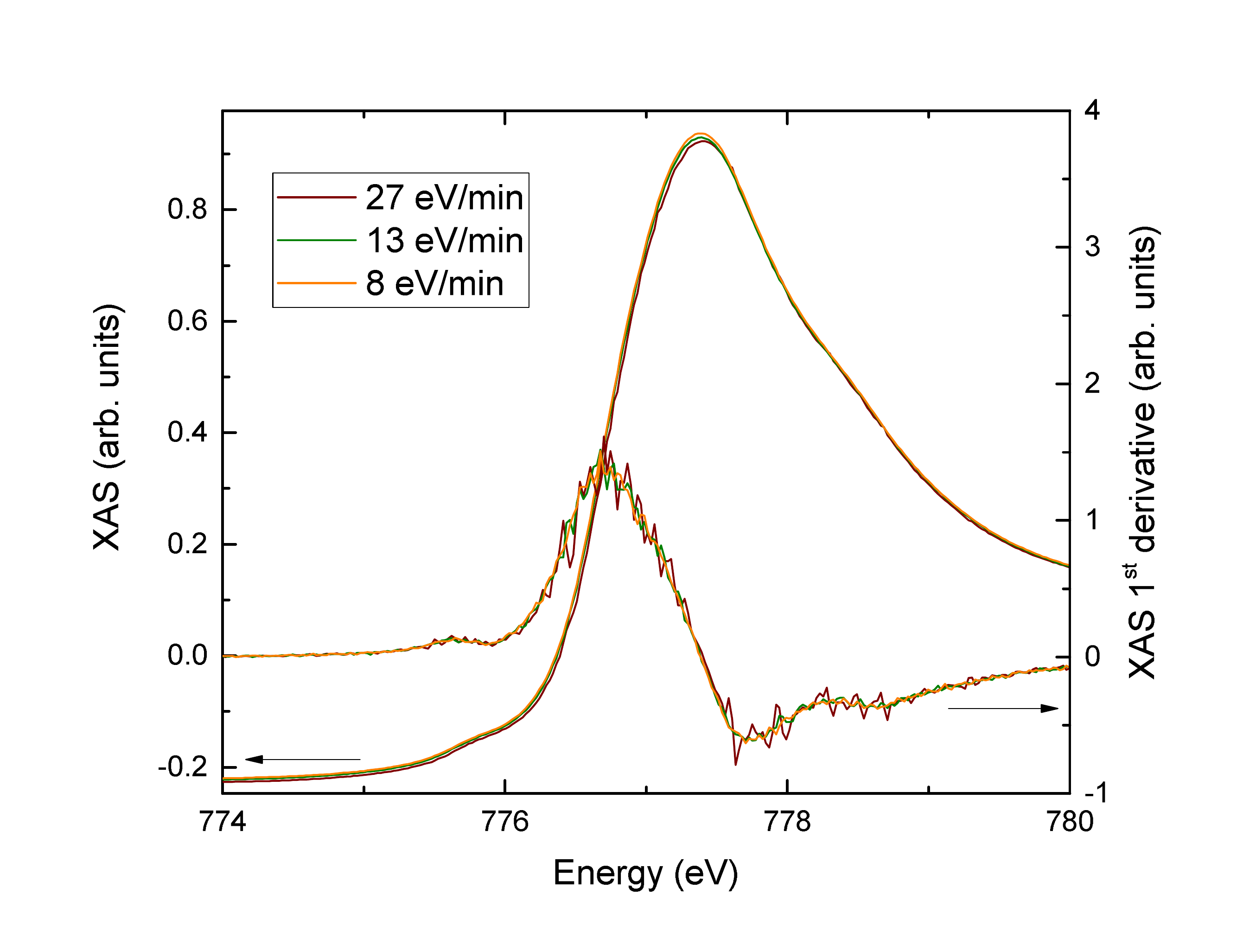}
\caption{\label{fig:XAS_speed} XAS at Co \liii\ edge and its derivative for spectra measured at different  speeds of the monochromator. }
\end{center}
\end{figure}

In many contemporary soft x-rays beamlines scanning of the monochromator is conducted continuously, and not in discreet steps. If the luminescence decay is very slow, there could be an effect of the scanning speed on the spectrum shape. We have therefor checked the effect of scanning speed for all samples measured here. One example is shown in figure \ref{fig:XAS_speed} for 20\,nm Co deposited on MgO capped with 3\,nm of Pt. The XAS derivative is also plotted. As seen from this figure different scanning speeds  between 27eV/min and 8eV/min of the monochromator do not significantly affect the spectrum shape.

\subsection{XMCD sum rules}
\begin{figure}
\begin{center}
\includegraphics[width=1.0\textwidth]{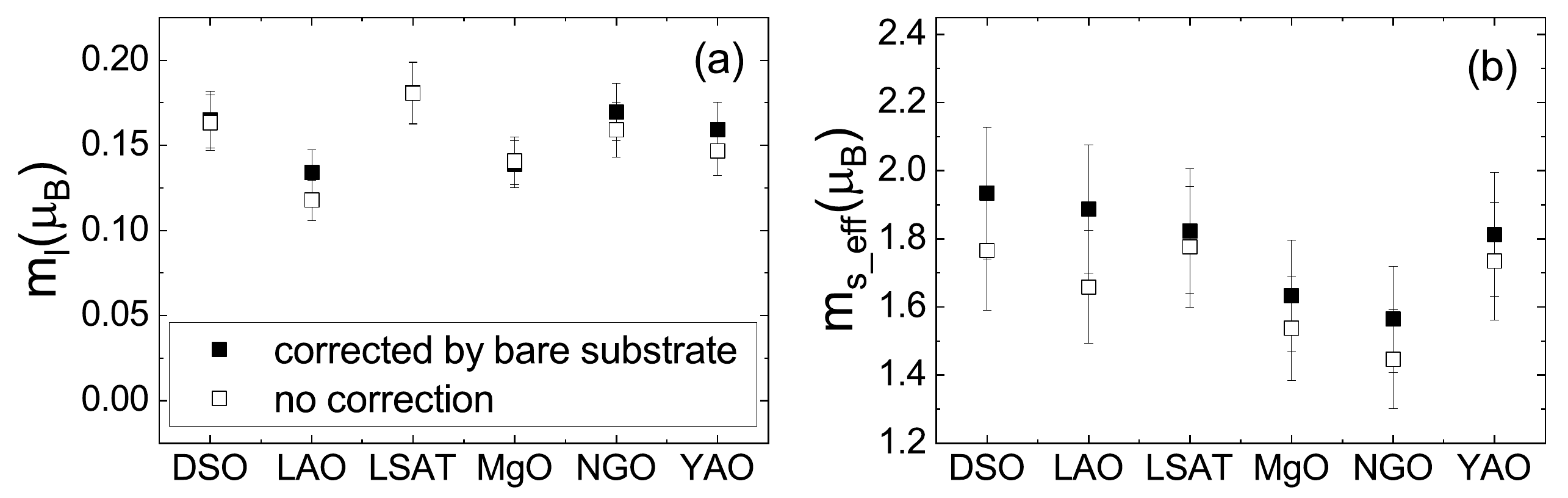}
\caption{\label{fig:sr} (a) Angular magnetic moment m$_{l}$ and (b) effective spin moment m$_{s_eff}$ obtained from sum rule analysis of the films in different substrates. The error bars correspond to $\pm10\%$ of the moment value. }
\end{center}
\end{figure}

We now focus on the application of XMCD sum rules on data generated using XEOL. XMCD sum rules\cite{Thole:1992tm,Carra:1993uz} were applied to  XMCD spectra of Co deposited on different substrates. The resulting spin and orbital moments are presented in figures \ref{fig:sr}(a) and (b). Because photoluminescence can depend on the incident photon energy, we tested whether the results of the sum rules analysis are affected by normalizing the XEOL signal of  thin film by the XEOL signal from  bare substrates (these were also measured in the same energy range). The filled symbols in figures \ref{fig:sr}(a) and (b) correspond to the data corrected by the bare substrate measurement as done in Ref. [\cite{Kallmayer:2007dz}]. The open symbols represent the analysis results without the  bare substrate normalization. We find that  m$_l$ is not significantly affected while m$_{s,eff}$ exhibits  systematically reduced values when the data are not normalized. The variations are however at a maximum of 5\% and therefore within the commonly used error bars of $\pm$10\%\cite{Laan:1999gy}. The substrate correction shows the largest effect on the Co/LAO values (see figure \ref{fig:sr}(a),(b)). This is caused by the La M$_{5}$ edge that is visible at the end of the Co absorption spectrum (see figure \ref{fig:substrate}). This affects the baseline  slightly even when limiting the used data energies below 820\,eV. This problem is not observed for LSAT, which has a much smaller La contribution. The effect of the LAO substrate exemplifies an obvious problem for the use of XEOL detection: the energy range of interest should not significantly overlap with resonances of the substrate. 

\subsection{Sample thickness}
\begin{figure}
\begin{center}
\includegraphics[width=1.0\textwidth]{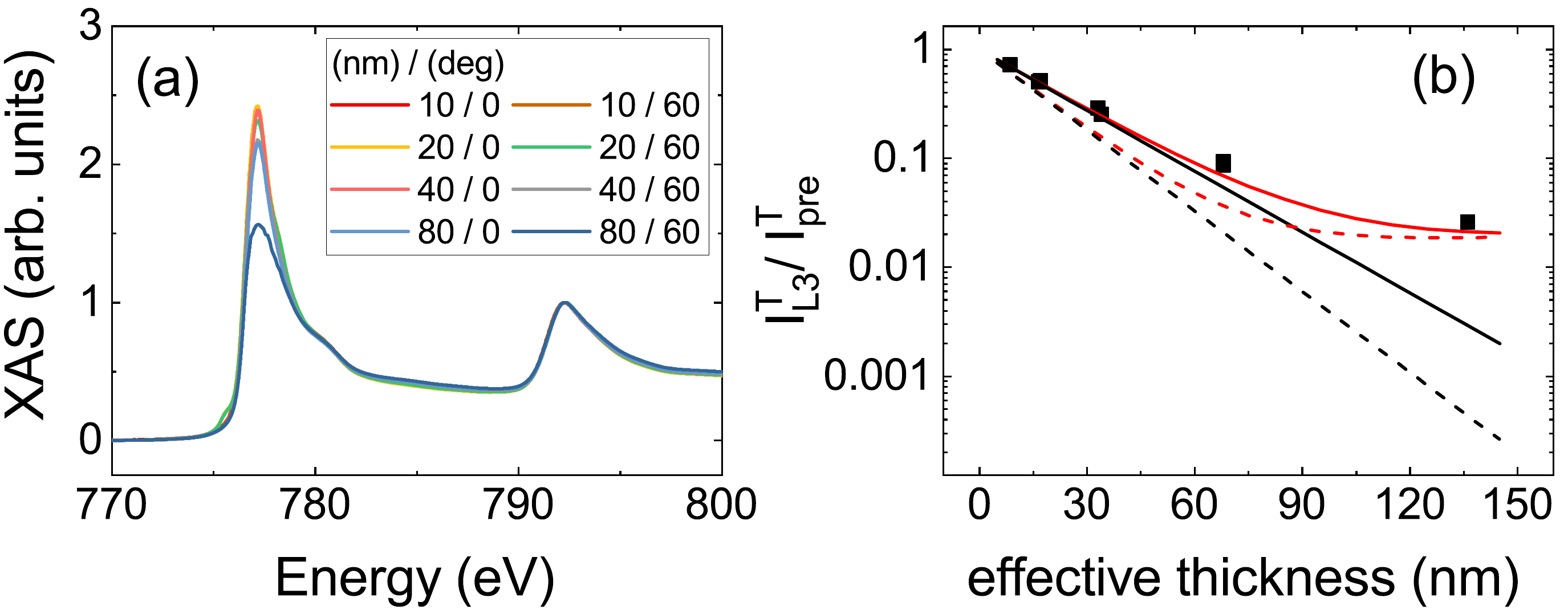}
\caption{\label{fig:saturation} (a) XAS data normalized at the L$_2$ edge. The legend gives the nominal thickness and incidence angle. (b) Transmitted intensity ($I^T=I/I_0$) taken at the peak of the \liii\ edge divided by the $I^T$ value at pre-edge plotted against the  effective thickness. Black (red) curves correspond to simulation for $\alpha=0$ ($\alpha=1.4\%$). Continuous (dashed) curve correspond simulation for $\mu=45/ \mu m$ ($\mu=59/ \mu m$).}
\end{center}
\end{figure}

\begin{figure}
\begin{center}
\includegraphics[width=1.0\textwidth]{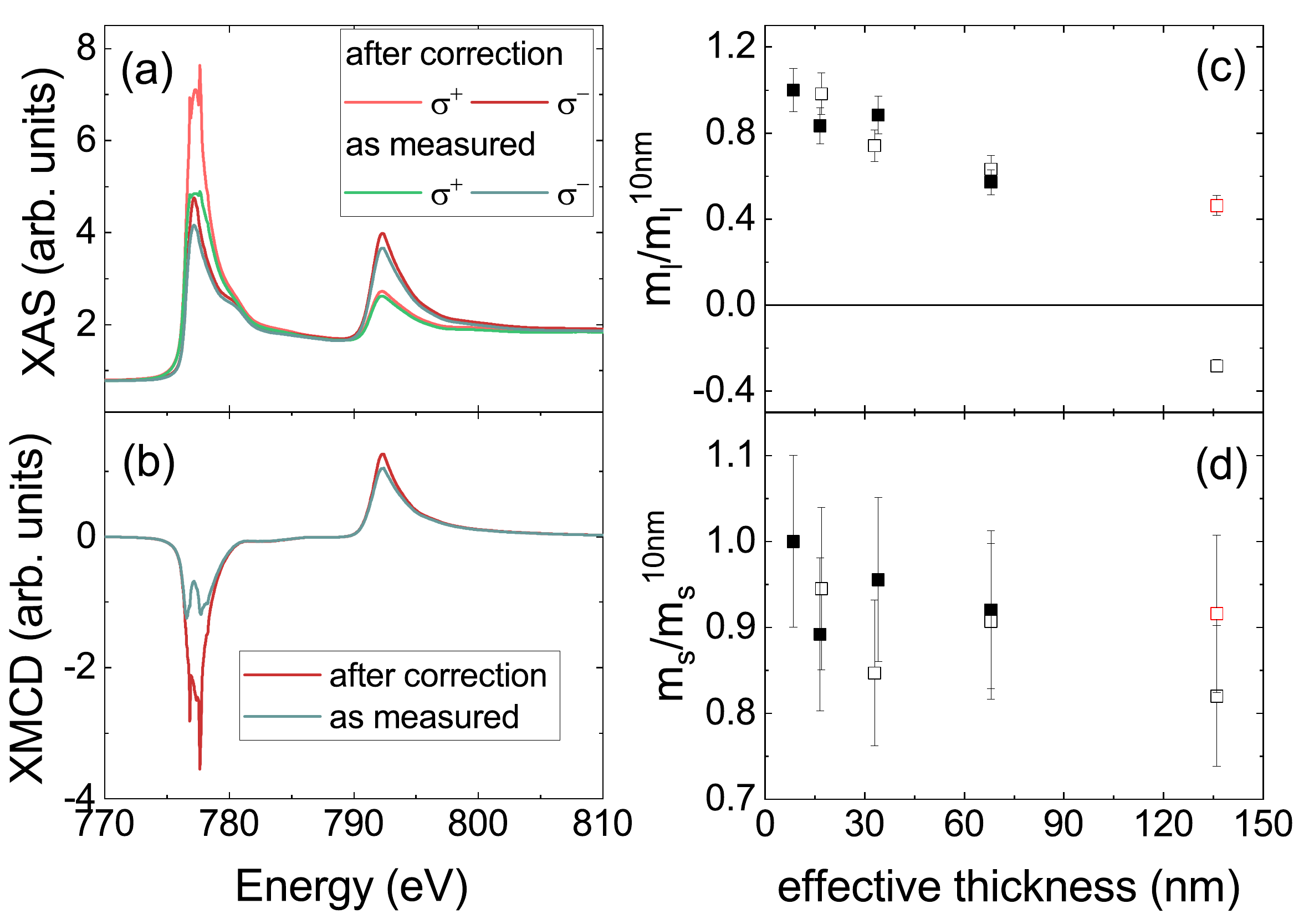}
\caption{\label{fig:corr}  (a) XAS for left and right circular polarization  and  (b) XMCD for Co nominal thickness of 80\,nm at 60\degree\   incidence.   (c) Orbital moment  and (d) effective spin moment plotted as a function of effective thickness. The data were normalized by the values for the 10\,nm Co film in normal incidence. Filled symbols show the data measured in normal incidence while open symbols show the measurements at 60\degree  incidence. The red open square corresponds to the 80\,nm data in grazing after the saturation correction discussed in the text. }
\end{center}
\end{figure}

XEOL may be used to overcome the known shortcomings of other detection modes. TFY detection suffers from self-absorption for samples with high concentrations of the absorbing element  \cite{Troger:1992ud,Iida:1993en}, while TEY detection exhibits saturation effects for very shallow incidence angles  \cite{Nakajima:1999tf}. For XEOL, there could also be self-absorption of the photoluminescence at the substrate. However, the good agreement between XEOL and transmission in figure \ref{fig:transmission} indicates that this is not visible for 0.5\,mm-thick MgO, probably due to the low absorption cross section of the optical photons. Transmission measurements should not saturate, but a saturation-like effect has been observed for samples with inhomogeneous thickness \cite{Hanhan:2009gs}. Given that XEOL is a transmission equivalent measurement, we checked if it suffers from the same saturation-like effects. We have measured Co films of several thicknesses at 0 and 60\degree\  incidence angles. The corresponding XAS spectra normalized at the maximum of the L$_2$ edge are presented in figure \ref{fig:saturation}(a). We find that the branching ratio between \liii\ and \lii\ depends on thickness, with larger thickness exhibiting a saturation-like effect at the \liii\ edge. That is specially true for the 80\,nm thick sample at 60\degree\ incidence (i.e. effective thickness of about 160\,nm). Figure \ref{fig:corr}(b) presents the corresponding XMCD spectra for this film, in which the \liii\ XMCD (light gray curve) appears clearly quenched.

Next we investigate how the saturation effect affects the sum rules results. Given that the orbital moment is proportional to the total XMCD integral, it is much more susceptible to the saturation effect than m$_{s,eff}$. Figures \ref{fig:corr}(c), (d) present results of the the sum rules analysis applied to the Co films of different thicknesses. Indeed, the effective spin moment does not vary much among the samples. The orbital moment clearly decreases as the thickness increases. For the 80\,nm film at 60\degree\  incidence, for which saturation is clearly seen in figure \ref{fig:saturation}(a), the orbital  moment even changes sign. 

\subsection{Saturation correction}
We now model and correct the data for such saturation-like effects. As pointed out by Stern and Kim \cite{Stern:1981vi} hard x-ray transmission signals can exhibit a saturation effect if part of the incoming x-rays are not absorbed by the sample, for example, due to high order contamination in the incoming beam or pinholes on the sample. For $b(E)$ being the contribution to the incoming x-rays which is not absorbed by the sample, the measured normalized transmitted signal is given by:

\begin{equation}\label{eq:transmission}
\frac{I^\prime}{I_0^\prime}=\frac{I+b}{I_0+b}=\frac{I_0e^{-\mu t}+b}{I_0+b}
\end{equation}

In equation (\ref{eq:transmission}), $I_0^\prime$ and $I^\prime$ correspond to the measured quantities, while $I_0$ and $I$ represent the intrinsic quantities. If we call  $\alpha(E)=b(E)/I_0(E)$, we obtain:

\begin{equation} \label{eq:alpha}
\frac{I_0^\prime}{I^\prime}=\frac{1+\alpha}{e^{-\mu t}+\alpha} 
\end{equation}

Equation (\ref{eq:alpha}) shows how the measured data relates to the  linear absorption coefficient $\mu$ and to $\alpha$.

Figure \ref{fig:saturation}(b) shows the ratio of the transmitted signals $I^T=I/I_0$, taken at the energy of maximum absorption ($\approx$ 777.15\,eV, named $I^T_{L3}$) over those taken at the pre-edge ($\approx$760\,eV, named $I^T_{pre}$). The curves in figure \ref{fig:saturation}(b) are simulations based on expression (\ref{eq:alpha}) for two different values of the absorption coefficient $\mu$ at the resonance:  $\mu=59/\mu m$ as obtained by Regan \textit{et al}\cite{Regan:2001bz} and $\mu=45/\mu m$, which best describes our data. The change in $\mu$ only affects the  agreement with the data at the low thickness range. An absorption coefficient of 2/$\mu m$ was used for the pre-edge in all cases. For the black curves no saturation effect is considered ($\alpha=0$). This results in a linear thickness dependence of the ratio $I^T_{L3}/I^T_{pre}$, as expected. The red curves in figure \ref{fig:saturation}(b)  are for  $\alpha=1.4\%$. The simulation for $\alpha=1.4\%$ and $\mu=45/\mu m$ is in good agreement with the measured data. The aim here is not to get a perfect agreement but to obtain an understanding for the saturation and attempt to correct for it. For simplicity, we consider $\alpha$ as a constant, but it could be energy dependent. 


If now we invert expression (\ref{eq:alpha}) to obtain the intrinsic $I$/$I_0$ and consider  $\alpha << 1$ we obtain:

\begin{equation}
\frac{I}{I_0}=\frac{I^\prime}{I_0^\prime} - \alpha
\end{equation}

Therefore, in order to correct our data, it is sufficient  to subtract the offset $\alpha$ from the measured $I^\prime/I_0^\prime$. The data in figure \ref{fig:saturation}(b)  is the sum of left and right helicities. Therefore the offset to be subtracted from each individual helicity spectrum is $\alpha/2$. Note that the offset subtraction is performed on the transmitted data, before the logarithm is taken to obtain the absorption spectrum (see equation \ref{eq_xas}).  The $\alpha$ contribution is likely present for all thicknesses, but only has a significant effect on the thickest samples. This is demonstrated by the simulations in figure \ref{fig:saturation}(b), where the inclusion of $\alpha$ significantly changes the edge jump only for samples with thickness above  50\,nm. 

The XAS and XMCD after the removal of $\alpha/2$ from  I$^T$  are plotted in figures \ref{fig:corr}(a),(b). The effect of the correction is clearest on the \liii\ peak of the positive helicity spectrum and on the XMCD spectrum. The corresponding sum rules results for this spectrum are plotted in figure \ref{fig:corr}(c),(d) by the red open squares. Even if the corrected spectrum may appear noisy, the sum rules results are in reasonable agreement with those from the thinner films, showing that the offset removal has satisfyingly corrected the data, and when properly considered, it can extend the thickness limitation on films measured using XEOL.

So far we have shown how to correct for the saturation effect. In the following we discuss the possible sources of this effect.  First we investigated possible spurious contributions arising  from an offset in the detection system. To test this, we repeated the measurements with an order of magnitude less flux, since a fixed offset in the detection system  would make the saturation flux dependent.  We find that the XAS is qualitatively the same for all thicknesses for both flux conditions, ruling out this hypothesis. Next we explored if this effect could come from higher-order diffraction of the monochromator. Reducing the higher-order contamination of the beamline from 0.5\% (standard setting) to 0.03\% by changing the monochromator c$_{ff}$ value \cite{Piamonteze:2012jg} did not result in a change in the branching ratio. Another possible factor that could cause  the saturation effect is thickness inhomogeneity, as discussed by Stern and Kim\cite{Stern:1981vi} and also pointed out by Kallmayer \textit{et al} \cite{Kallmayer:2007dz}. We  estimate that this correction, if coming from typical roughness values, is too small. The only option which seems to explain our data is the  contribution of fluorescence photons emitted from the Co film itself. Since the fluorescence photons have an energy below the \liii\ absorption edge the attenuation length is about 1/2\,$\mu m$ (Co pre-edge), and will be mostly transmitted to the substrate, where they create additional luminescence. The L shell fluorescence efficiency is about 0.5\% for Ni \cite{Auerhammer:1988tw} and it should be of the same order for Co. Only the fluorescence going in the direction of the substrate will generate luminescence, which approximately halves this value.  These Co fluorescence photons will act as additional x-ray intensity reaching the substrate generating additional optical luminescence. A rough estimate using the numbers discussed above suggests that the amount of fluorescence from a thick Co film reaching the substrate would be about 1.5-2\% of I$_0$, which is in the same order as the correction factor used here. 

The question that remains is what is a safe thickness for XEOL measurements. In figure \ref{fig:saturation}(b) we see that a significant divergence between the simulations for $\alpha=0$ and $\alpha=1.4\%$ starts around 30-50\,nm effective thickness.   Taking the absorption coefficient at the \liii\ resonance from our simulation in figure \ref{fig:saturation}(b) of 0.045/nm times an effective thickness of 50\,nm we obtain $\mu\times t = 2.25$, which is within the range suggested by previous works of $\mu \times t$=2.6\cite{Rose:1948vp} and 1.5\cite{Stern:1981vi} (for EXAFS). A guideline in how to calculate $\mu$ at the resonance is given, for example, by Regan \textit{et al} \cite{Regan:2001bz}. 

\section{Conclusions \label{sec:conclusions}}

We have shown that XEOL detection of the substrate is a transmission equivalent detection mode of the thin film on top. We have measured the XEOL efficiency for a range of substrates, which should serve as a guideline for designing a XAS experiment based on XEOL detection. We have shown that XMCD sum rules applied to XEOL detected data give reasonable results independent on the analysis procedure. Analogously to transmission measurements, XEOL shows a saturation-like effect, which appears when part of the x-ray intensity is not absorbed by the sample. The saturation has a drastic effect in the result of the orbital sum rules. We have given guidelines for the film thickness to avoid this effect, keeping $\mu\times t < 2.2$ at the maximum absorption, and we have shown how to successfully correct for the saturation when a range of film thicknesses is measured. In our case, we consider that the saturation effect comes from the x-ray fluorescence of the thin film in itself which excites additional optical luminescence at the substrate. 
 
Finally XEOL is  a good complementary detection method to TEY, due to the different probing depths allowing the comparison between surface and bulk effects in thin films\cite{Aeschlimann:2018jga}. Similarly to TFY it does not require a conductive sample. On the other hand, it does not  suffer from known problems of TFY interpretation\cite{deGroot:2012kj,deGroot:2012kj,Liu:2017fc}. This method is useful not only in synchrotron beamlines but also in experiments using high harmonic generation sources.


%

\section{Acknowledgments}
We wish to acknowledge E. Arenholz for helpful  discussions about the XEOL setup in the Advanced Light Source. The measurements here were done using the EPFL/PSI endstation at the X-Treme beamline. We would like to thank the technical support of the X-Treme beamline in preparing the sample holders dedicated to XEOL measurements. SRVA and YWW thank the Swiss National Science Foundation for financial support under projects No. 169467 and No. 137657, respectively.

\bibliography{ms}

\end{document}